\documentstyle[aps,prl,epsfig,preprint,upciting]{revtex}
\renewcommand{\vec}[1]{\relax\ifmmode\mathchoice
{\mbox{\boldmath$\relax\displaystyle#1$}}
{\mbox{\boldmath$\relax\textstyle#1$}}
{\mbox{\boldmath$\relax\scriptstyle#1$}}
{\mbox{\boldmath$\relax\scriptscriptstyle#1$}}\else
\hbox{\boldmath$\relax\textstyle#1$}\fi}
\begin{document}
\pagestyle{myheadings}
\markboth{Helbing/Vicsek: Self-Organised Optimality ...}
{Helbing/Vicsek: Self-Organised Optimality ...}
\title{\mbox{}\\[-2.6cm]Self-Organised Optimality in Driven Systems
with Symmetrical Interactions\\[-0.5cm]}
\author{Dirk Helbing$^{*,+}$ and Tam\'{a}s Vicsek$^+$}
\address{$^*$ II. Institute of Theoretical Physics, University of Stuttgart,\\
Pfaffenwaldring 57/III, 70550 Stuttgart, Germany\\
$^+$ Department of Biological Physics, E\"otv\"os University, Budapest,\\
P\'azm\'any P\'eter S\'et\'any 1A, H-1117 Hungary\\
{\tt helbing@theo2.physik.uni-stuttgart.de; vicsek@angel.elte.hu}\\[1mm]}
\maketitle
\draft
{\bf Extremal principles
are fundamental in our interpretation of phenomena in 
nature. One of the best known examples is the second law of
thermodynamics,\cite{thermo}
$^-$\cite{Pri} governing most physical and chemical systems
and stating the continuous increase of entropy in closed systems.
Biological
and social systems, however, are usually open and
characterised by self-organised
structures.\cite{Info}
$^-$\cite{gamedyn} Being results of an
evolutionary optimisation process,\cite{frank}
$^-$\cite{gamedyn} 
one may conjecture that such systems
use resources like energy very efficiently, but there is no proof for this.
Recent results on driven systems\cite{trail,solid}
indicate that systems composed of competing 
entities tend to reach a state of self-organised optimality
associated with minimal interaction or minimal dissipation,
respectively. Using concepts from non-equilibrium 
thermodynamics\cite{thermo}
$^-$\cite{Pri,Ebeling} and game theoretical
ideas,\cite{Ebeling}
$^-$\cite{may} we will show that
this is universal to an even wider class of systems  
which, generally speaking, have 
the ability to reach a state of maximal overall ``success''. 
This principle is expected to be relevant
for driven systems in physics, but its main significance concerns
biological and social systems, for which  only a limited number
of quantitative principles are available yet.}
\clearpage

\begin{figure}[htbp]
\begin{center}
\begin{picture}(5,5.6)(-0.7,-0.5)
\thicklines
\put(0,0){\circle*{0.15}}
\put(0,4){\circle*{0.15}}
\put(4,0){\circle*{0.15}}
\put(4,4){\circle*{0.15}}
\put(0,0){\line(1,1){1}}
\put(0,4){\line(1,-1){1}}
\put(4,0){\line(-1,1){1}}
\put(4,4){\line(-1,-1){1}}
\put(1,1){\line(1,0){2}}
\put(1,1){\line(0,1){2}}
\put(3,3){\line(-1,0){2}}
\put(3,3){\line(0,-1){2}}
\end{picture}
\end{center}
\caption[]{Schematic representation of the human trail system evolving
on an initially homogeneous ground.\cite{trail}
When the frequency of trail usage
is small, the minimal way system (that would connect the
four entry points and destinations in the corners by
direct ways) cannot be supported in competition with the regeneration
of the vegetation. Here, by bundling of trails, the frequency of usage
becomes large enough to support the depicted trail system. It corresponds
to the optimal compromise between the diagonal ways and the
ways along the edges, supplying optimal walking comfort at a minimal
detour of 22\% for everyone, which is a fair solution.
\label{fig1}}
\end{figure}
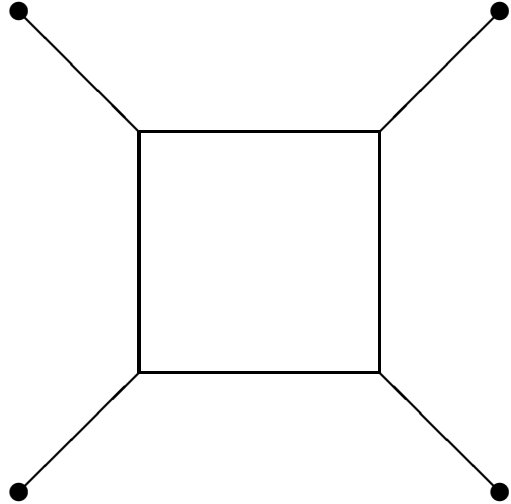

Recent simulations of driven multi-particle or multi-agent systems
have revealed different kinds of self-organised states that seem to
optimise certain aggregate quantities. Hence, in analogy with the concept of
``self-organised criticality'' for systems driving themselves to
a critical state,\cite{SOC} it is natural to introduce the concept
of ``self-organised optimality''. As a specific
example, one can consider the formation of human trail systems,\cite{trail}
where it is the discomfort of walking multiplied by the length of the
individual ways that is minimised 
(cf. Figure~\ref{fig1}).
\par
\begin{figure}[htbp]
\begin{center}
\epsfig{width=14.0\unitlength, angle=0,
      file=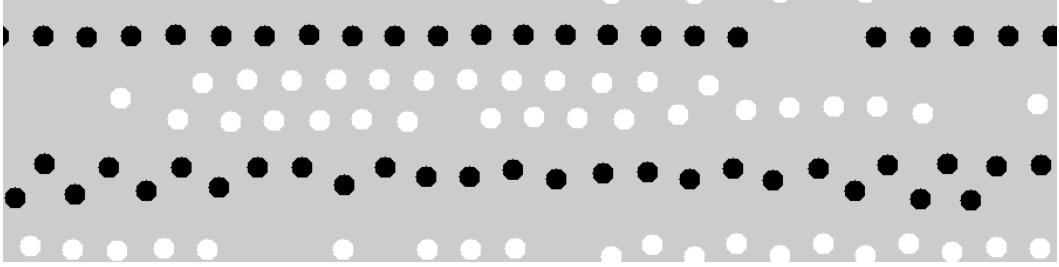} 
\end{center}
\vspace*{6mm}
\caption[]{Formation of lanes of
uniform walking directions in crowds of
oppositely moving pedestrians. Blue 
circles represent pedestrians walking
from left to right, red 
ones represent pedestrians walking into the
opposite direction.
If $\vec{x}_\alpha(t)$ denotes the position of pedestrian $\alpha$
at time $t$, $\vec{v}_\alpha(t) = d\vec{x}_\alpha(t)/dt$
its velocity, $v_0$ the desired speed, and $\vec{e}_\alpha$
the desired walking direction, our simplified pedestrian model
reads
$\vec{v}_\alpha(t) = v_0 \vec{e}_\alpha + \sum_{\beta (\ne \alpha)}
 \vec{f}_{\alpha\beta}\big(\vec{x}_\alpha(t),\vec{x}_\beta(t)\big)$.
Here, $\vec{f}_{\alpha\beta}$ represents repulsive interactions between
pedestrians $\alpha$ and $\beta$, which were assumed to decrease
monotonically with their distance $d_{\alpha\beta}(t)=
\|\vec{x}_\alpha(t) - \vec{x}_\beta(t)\|$.
For simplicity, we have specified the interactions as a gradient of
a rotationsymmetric potential that depends only on $d_{\alpha\beta}$.
Notice that lane formation is not a trivial effect of this model, 
but it eventually arises
due the smaller relative velocity
and interaction rate that pedestrians with the same walking
direction have.
It is clear that lane formation will increase
the average velocity in walking direction
$E(t) = \langle\langle \vec{v}_\alpha \cdot \vec{e}_\alpha
 \rangle_\alpha \rangle_t / v_0 \le 1$, which is a measure of ``efficiency''
or ``success''.
(Here, $\langle\langle . \rangle_\alpha \rangle_t$ denotes the
average over the pedestrians and over time.) Moreover, maximisation of
efficiency immediately implies that 
the system minimises the quantity
$\langle\langle - \sum_{\beta (\ne \alpha)}
 \vec{f}_{\alpha\beta} \cdot \vec{e}_\alpha \rangle_{\alpha}
 \rangle_{t}
 = v_0 - \langle\langle  v_\alpha \cdot \vec{e}_\alpha
 \rangle_\alpha\rangle_t = v_0 (1 - E)$,
i.e., the average interaction against the desired direction of motion.
(The average interaction perpendicular to it is zero.)
For dissipative interactions, minimal dissipation is a direct
consequence of minimal interactions.
\label{fig2}}
\end{figure}

Here, we will focus on the dynamics of pedestrian crowds, allowing
the
intuitive and clear illustration
of the non-trivial mechanisms behind self-organised optimality.
In crowds of oppositely moving pedestrians,
usually lanes of uniform walking
direction develop\cite{pre} (cf. Figure~\ref{fig2}).
Obviously, this self-organised collective pattern of motion maximises
the average speeds and minimises interactions,
since pedestrians would be considerably slowed down by
avoidance maneuvers,
if the desired walking directions were mixed.
\par
To prove minimal interaction, we will set up continuum equations for the
considered systems,
but at the same time, we will generalise the problem to arbitrary
kinds of populations $a$ that are composed of identical moving entities
$\alpha$. In the above examples,
the different populations correspond to different origin-destination
pairs and desired walking directions, 
respectively. The distribution of the $N_a$ entities of population
$a$ over the locations $\vec{x}$ of an $n$-dimensional
space will be represented by the densities $\rho_a(\vec{x},t)\ge 0$.
The space does not need to be real. It may be an abstract one.
In the case of trail formation, for example, a point $\vec{x}$ in space
corresponds to a {\em path} connecting origin and destination.
\par
Assuming conservation of the number
\begin{equation}
  N_a = \int d^nx \, \rho_a(\vec{x},t)
\label{conserv}
\end{equation}
of entities in each population $a$, we obtain the
continuity equations\cite{Keizer}
\begin{equation}
 \frac{\partial \rho_a(\vec{x},t)}{\partial t}
 + \vec{\nabla} \cdot \Big[ \rho_a(\vec{x},t) \vec{V}_{\!\!a}(\vec{x},t)
 \Big] = 0 \, .
\label{cont}
\end{equation}
Here, $\vec{V}_{\!\!a}(\vec{x},t)$ is the average velocity of entities of
population $a$. We will assume that this is given by the gradient
\begin{equation}
 \vec{V}_{\!\!a}(\vec{x},t) = \vec{\nabla} S_a(\vec{x},t)
\label{gradient}
\end{equation}
of some function $S_a$ of the densities $\rho_b$. 
The existence of such a function 
requires the potential condition
$\partial V_{ai}(\vec{x})/\partial x_j
= \partial V_{aj}(\vec{x})/\partial x_i$ to
be true, but sometimes it can be relaxed. In particular,
it is not fulfilled in
our pedestrian model. However, assuming homogeneity in $x_1$-direction,
we can reduce the problem to the investigation of the one-dimensional dynamics
in $x_2$-direction, for which the potential condition is satisfied.
\par
The first terms of a series expansion of $S_a(\vec{x},t)$ give
\begin{equation}
 S_a(\vec{x},t) = S_a^0 + \sum_b S_{ab} \, \rho_b(\vec{x},t) \, ,
\label{success}
\end{equation}
but the constants $S_a^0$ do not matter at all.
The function $S_a(\vec{x},t)$ may be interpreted as the ``(expected) success''
per unit time for an entity of population $a$ at location $\vec{x}$,
as it is plausible that the entities move into the direction of the
greatest increase of success, corresponding to the gradient
(\ref{gradient}). Positive $S_{ab}$ belong to profitable 
or attractive
interactions between populations $a$ and $b$, whereas competitive or
repulsive interactions correspond to negative $S_{ab}$.
If an entity of kind $a$ interacts with
entities of kind $b$ at a rate $\nu_{ab}$ and the associated
result of the interaction can be quantified by some ``payoff'' $P_{ab}$, we
have the relation $S_{ab} = \nu_{ab}P_{ab}$.
However, the above equations differ from the conventional
game dynamical equations\cite{Ebeling,gamedyn} in several respects:
1. We have a topology (like in the {\em game of life}),\cite{life,may}
but define abstract games for interactive
motion in space with the possibility of local agglomeration
at a fixed number of entities in each population.
2. The payoff does not depend on the variables that the individual entities 
can change (i.e. the spatial coordinates $x_i$).
3. Individuals can only improve their success by redistributing
themselves in space.
4. The increase of success is not proportional to the
difference with respect to the global average of success, 
but to the local gradient of success in a population.
\par
\begin{figure}[htbp]
\begin{center}
\begin{picture}(16,15)
\put(0,15.5){\epsfig{height=7.5\unitlength, angle=-90,
      file=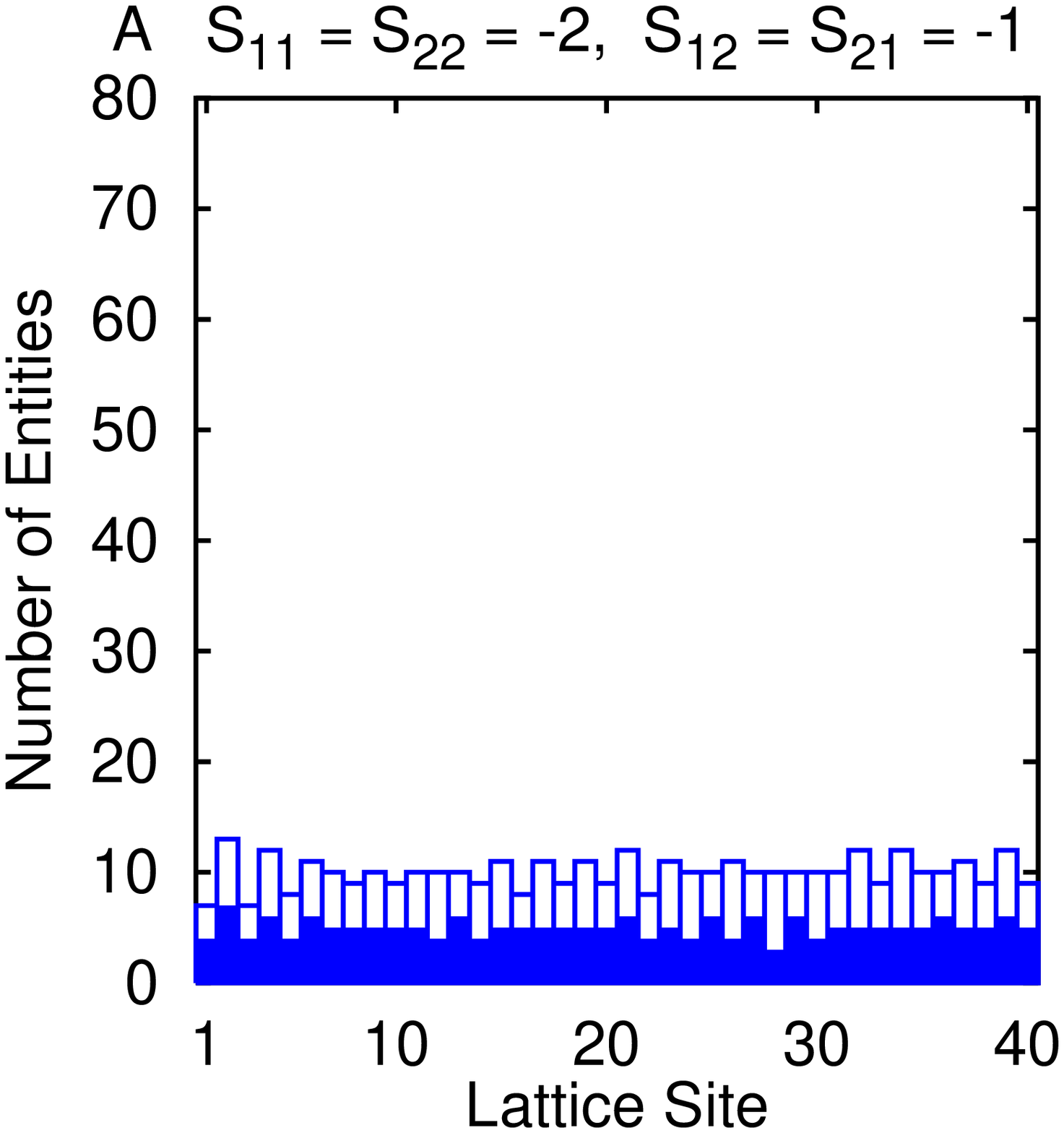}}
\put(1.8,14.8){\epsfig{height=4.8\unitlength, angle=-90,
      file=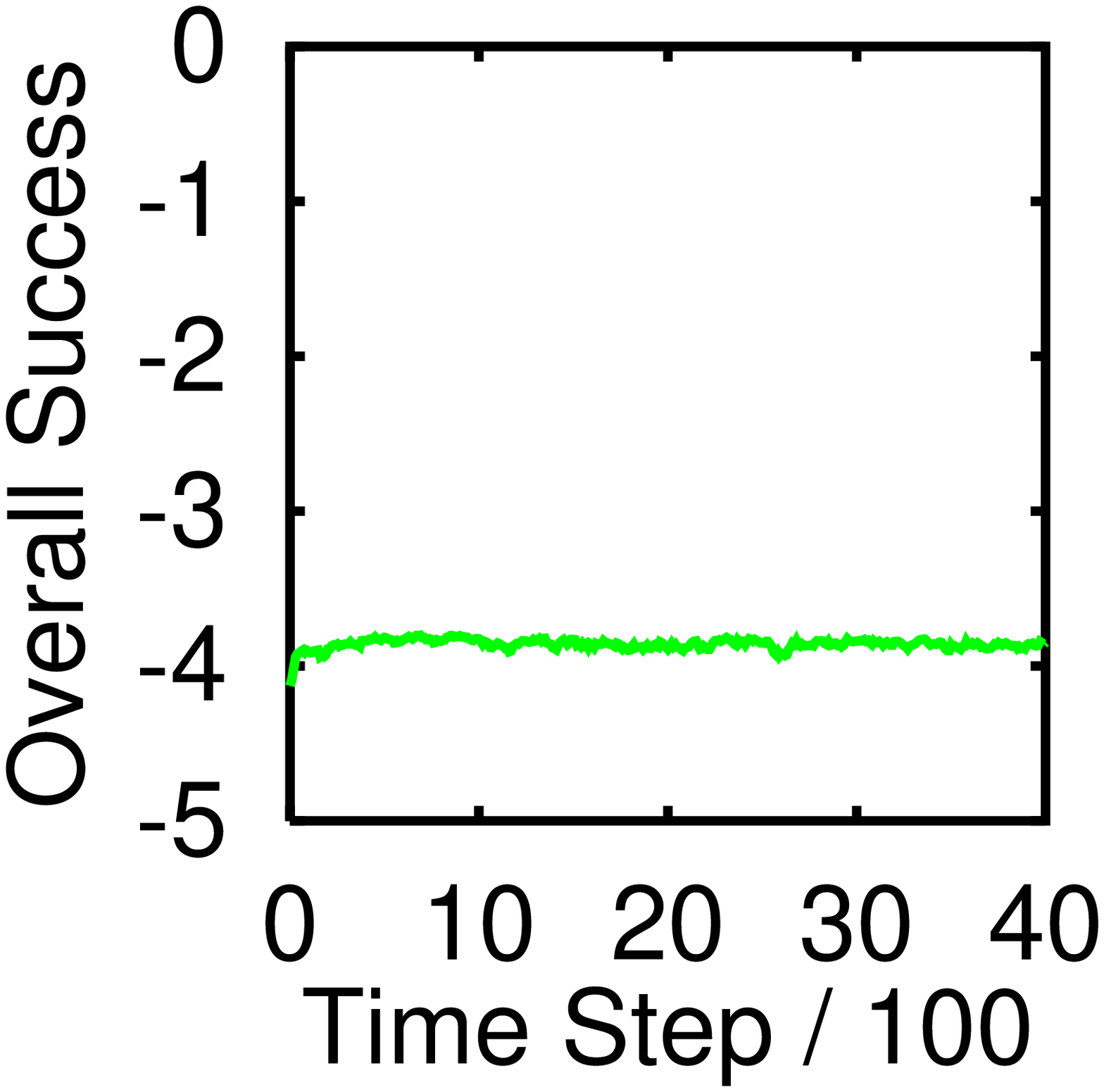}}
\put(7.5,15.5){\epsfig{height=7.5\unitlength, angle=-90,
      file=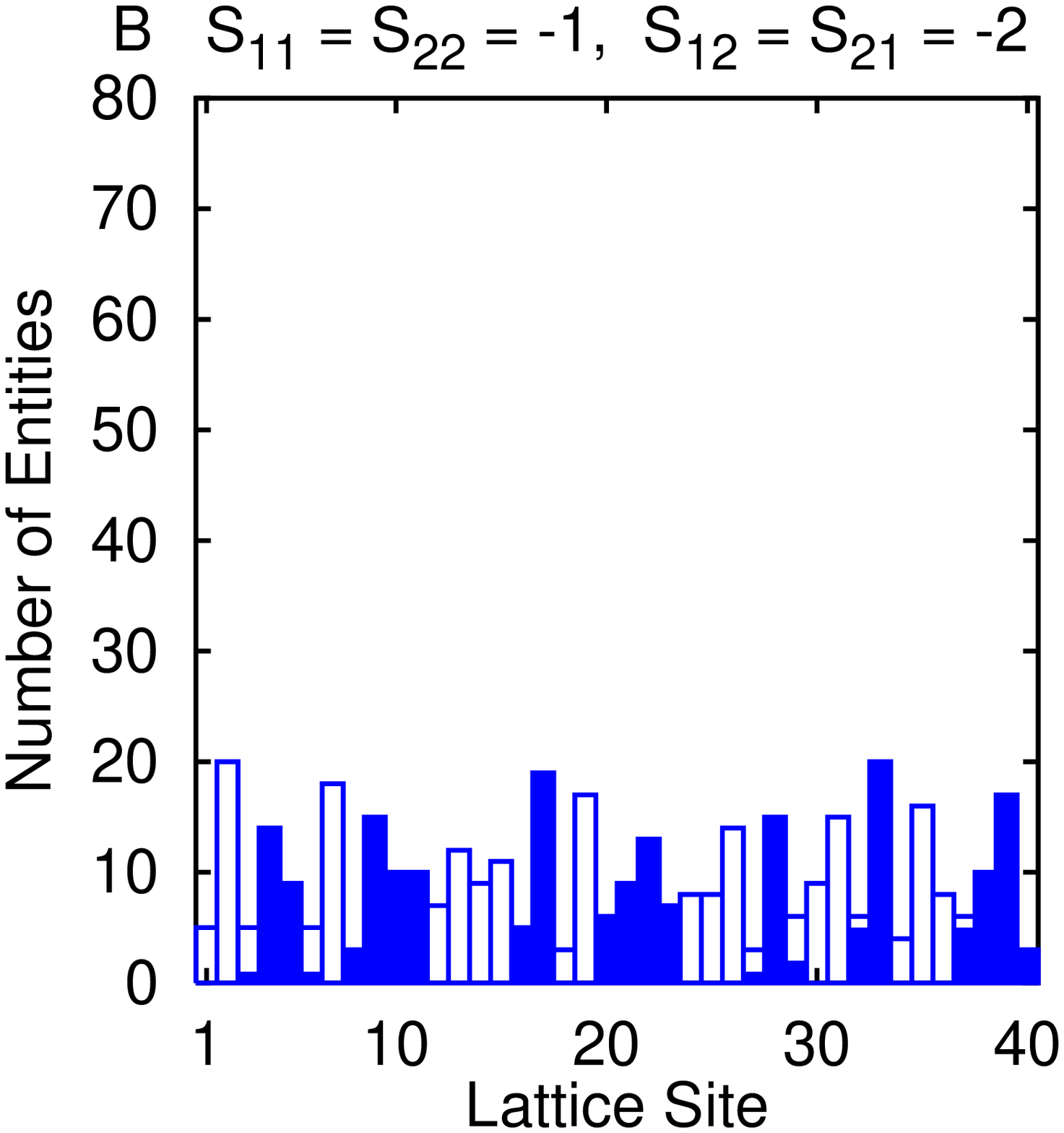}} 
\put(9.3,14.8){\epsfig{height=4.8\unitlength, angle=-90,
      file=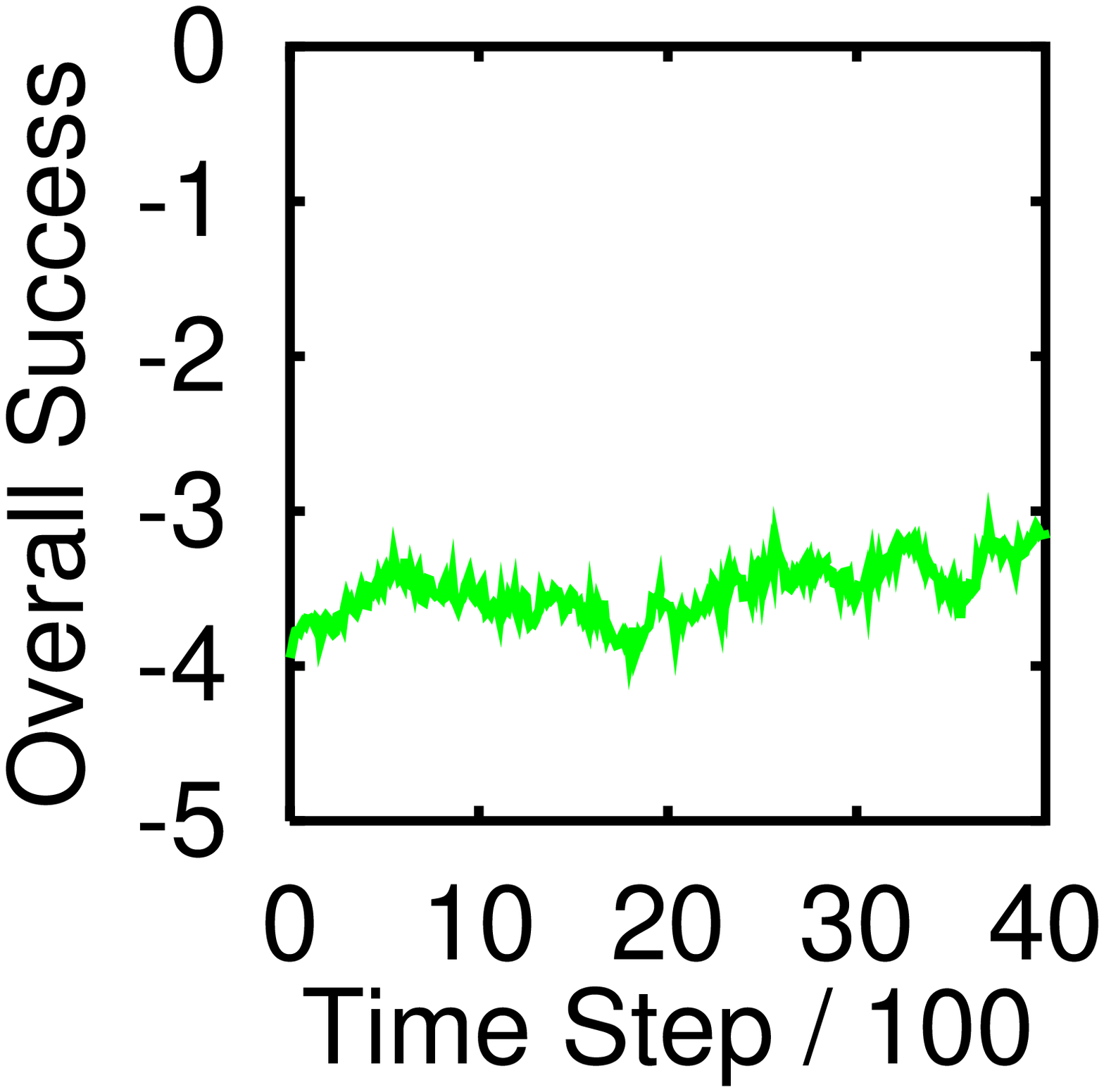}} 
\put(0,8){\epsfig{height=7.5\unitlength, angle=-90,
      file=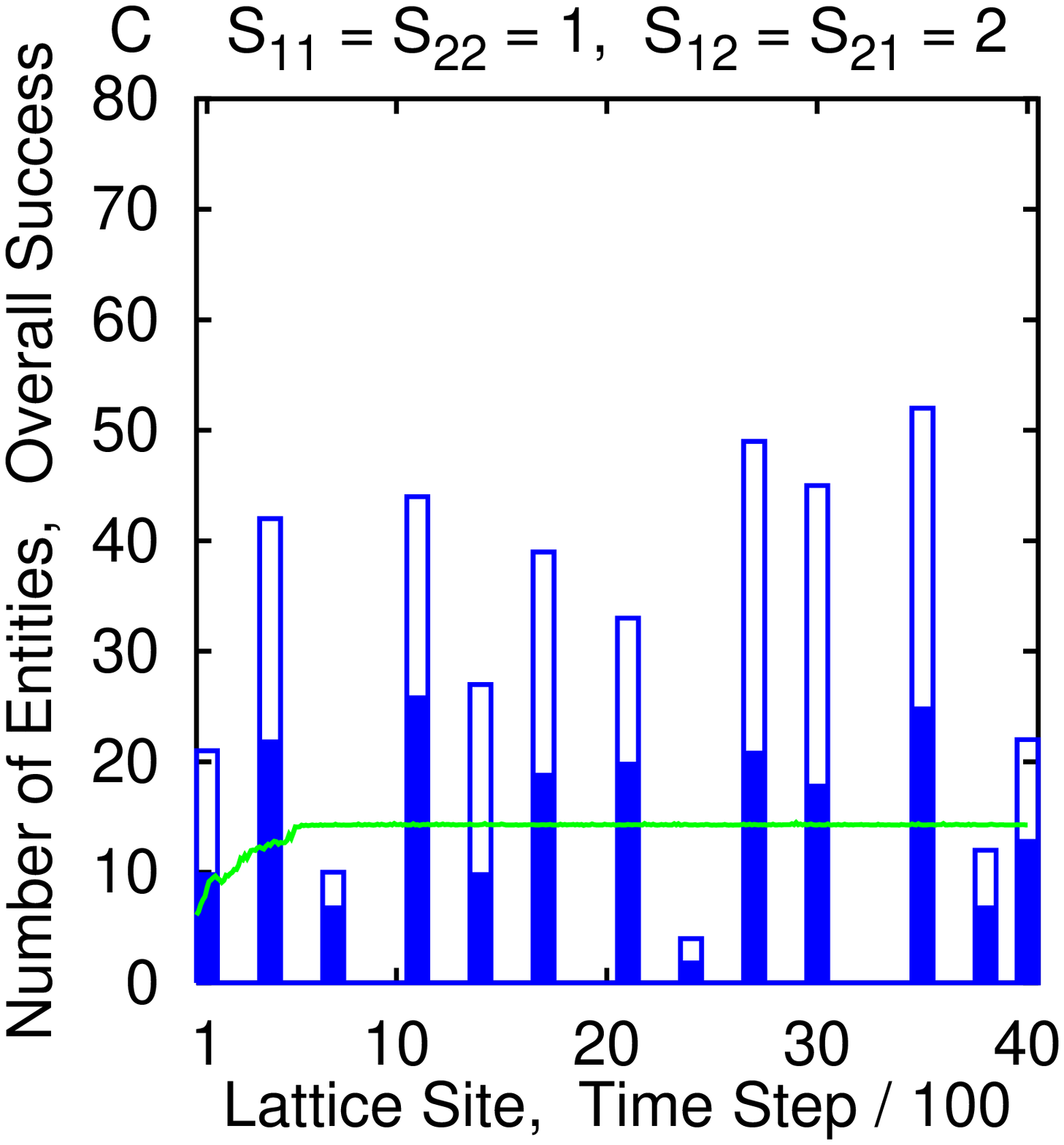}} 
\put(7.5,8){\epsfig{height=7.5\unitlength, angle=-90,
      file=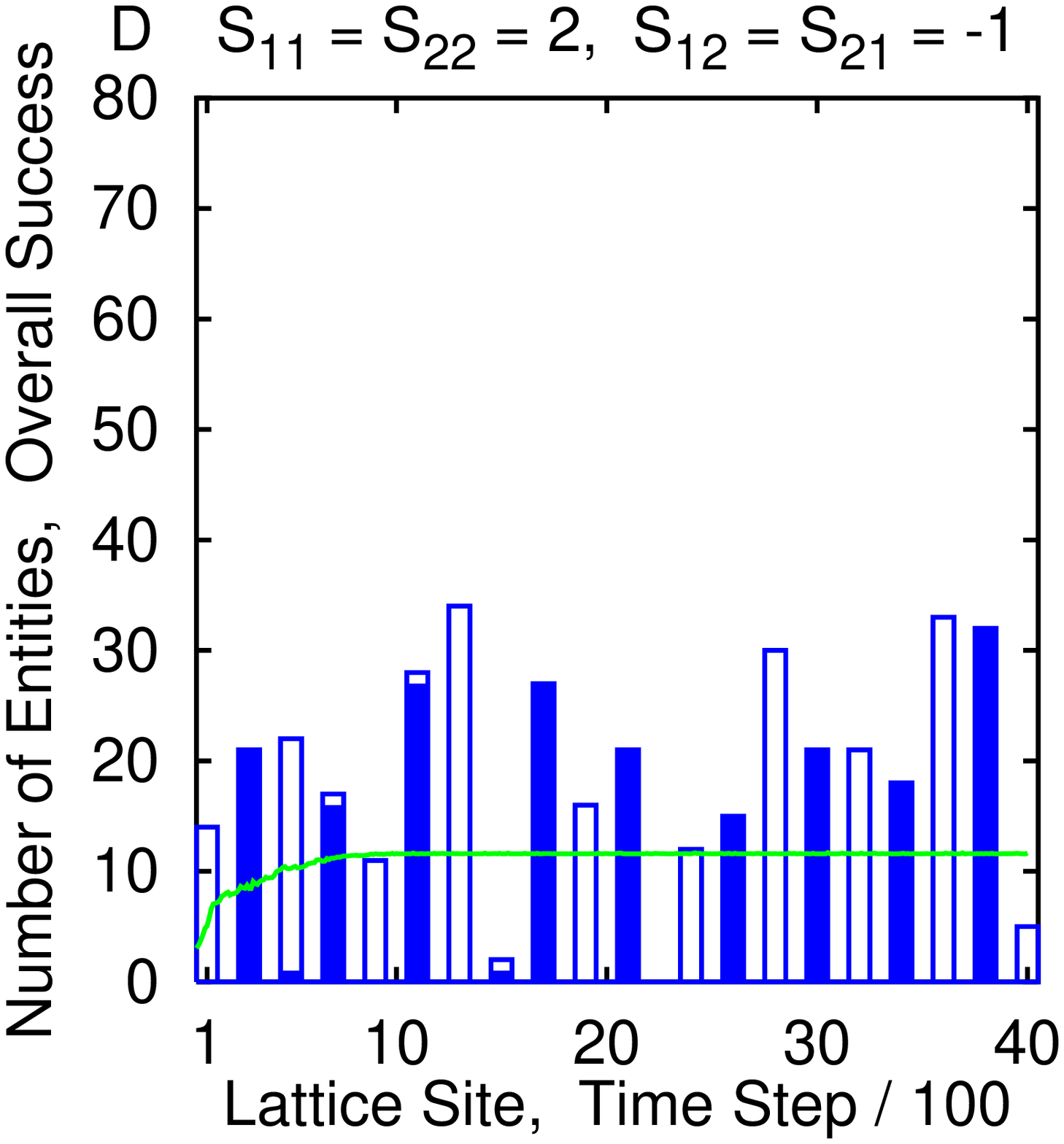}}
\end{picture}
\end{center}
\caption[]{Illustration of the various forms of self-optimisation
for two different populations:
(A) Homogeneous distribution in space, (B) segregation of 
populations without agglomeration, (C) attractive agglomeration, 
(D) repulsive agglomeration. In cases (B) 
to (D), the finally evolving optimal state is related with
a self-organised, non-homogeneous state, which corresponds to 
``self-organised optimality''. The above figures were obtained with a
one-dimensional, discrete version of the game-dynamical
model defined by equations (\ref{cont}) to (\ref{success}). 
We assumed a periodic lattice
with $V$ lattice sites $x\in\{1,\dots,V\}$ and two
populations $a\in\{1,2\}$ with a total of $N=N_1+ N_2 \gg V$ entities
(here: $V=40$ and $N_1 = N_2 = 200$).
Furthermore, we applied the following update steps: 1. Calculate the successes
$S_a(x,t) = S_a^0 + \sum_b S_{ab} n_x^b(t)/V$, where $n_x^b(t) =
\rho_b(x,t) V$ represents the
number of entities of population $b$ at site $x$.
2. For each entity $\alpha$,
determine a random number $\xi_\alpha$ that is uniformly
distributed in the interval $[0,S_{\rm max}]$ with a large constant 
$S_{\rm max}$ (here: $S_{\rm max}=20$).
3. Move entity $\alpha$ belonging to population $a$ from site $x$ to site
$x+1$, if $[S_a(x+1,t) - S_a(x-1,t)] > \xi_\alpha$, but to site
$x-1$, if $[S_a(x-1,t) - S_a(x+1,t)] > \xi_\alpha$. Our simulations
started with a random initial distribution of the entities. We applied
a random sequential update rule, but a parallel update yields 
qualitatively the same results. The above figures show 
the numbers $n_x^1$ and $n_x = (n_x^1+n_x^2)$ of entities as a function
of the lattice site $x$ at time $t=4000$ and the evolution of the 
overall success $S(t)$ as a function of time $t$. The fluctuations
around the monotonic increase of $S(t)$ are caused by 
the fluctuations $\xi_\alpha$ and the random sequential update.
\label{fig3}}
\end{figure}

Now, we will proof that, for symmetric interactions with $S_{ba} = S_{ab}$,  
the overall success
\begin{equation}
 S(t) = \sum_a \int d^n x \, \rho_a(\vec{x},t) S_a(\vec{x},t)
\end{equation}
is a so-called Lyapunov function which monotonically increases in the
course of time, just like some thermodynamic
non-equilibrium potentials.\cite{Graham}
Because of (\ref{conserv}), we eventually obtain $d S(t)/d t
 = \sum_{a} \int d^n x \, \partial \rho_a(\vec{x},t)/\partial t$
$\sum_b (S_{ab} +  S_{ba}) \rho_b(\vec{x},t)$.
By inserting (\ref{cont}) and (\ref{gradient}), and
applying (\ref{success}), we get
$d S(t)/dt =  -2 \sum_a \int d^n x \, \vec{\nabla} \cdot
 \big[ \rho_a(\vec{x},t) \vec{\nabla} S_a(\vec{x},t)
 \big] \big[ S_a(\vec{x},t) - S_a^0 \big]$.
Making use of partial integration and the Gaussian integral theorem
(for systems with periodic boundary conditions),
we finally arrive at
\begin{equation}
 \frac{dS(t)}{dt} =
 2 \sum_a \int d^n x \, \rho_a(\vec{x},t) \left[ \vec{\nabla}
 S_a(\vec{x},t) \right]^2 \ge  0 \, .
\label{endresult}
\end{equation}
This result 
establishes self-optimisation for symmetical interactions
and can be easily transferred to discrete spaces 
(see Figure~\ref{fig3}). Notice that $S(t)$ is bounded for any
finite system and that
(\ref{endresult}) looks similar to dissipation functions in
thermodynamics.\cite{thermo,Keizer,Graham} 
If we interpret the function $-dS(t)/dt$
as a measure of dissipation per unit time in the system,
Eq.~(\ref{endresult}) immediately implies that the system approaches
a state of minimal dissipation.
\par
According to (\ref{endresult}), the stationary solution
$\rho_a^{\rm st}(\vec{x})$ is characterised by
\begin{equation}
 \rho_a^{\rm st}(\vec{x}) = 0 \qquad \mbox{or} \qquad
 \vec{\nabla} S_a^{\rm st}(\vec{x}) = \sum_b S_{ab} \vec{\nabla}
 \rho_b^{\rm st}(\vec{x}) = \vec{0}
\end{equation}
for all $a$, which is fulfilled by homogeneous or
step-wise constant solutions. For the case of two species,
a linear stability analysis shows that the homogeneous solution 
(cf. Figure~\ref{fig3}(A)) is unstable if
\begin{equation}
 \rho_a^{\rm hom} S_{aa} + \rho_b^{\rm hom} S_{bb} > 0 \quad \mbox{or} \quad 
 S_{ab} S_{ba} > S_{aa} S_{bb}  \, ,
\label{cond}
\end{equation}
where $\rho_a^{\rm hom}= N_a/V$ denotes the homogeneous density and
$V$ the volume of the system. The latter condition in (\ref{cond}) is, 
for example, fulfilled for lane formation by pedestrians, since their 
interaction rate is proportional to their relative velocity,
which is much higher for oppositely moving pedestrians than for pedestrians
with the same desired walking direction.
\par
Under the condition (\ref{cond}),
the stable stationary solution corresponds to complete segregation 
which, together with (\ref{endresult}), implies that the system
reaches a state of self-organised optimality. 
Notice that the global optimum is reached by means of short-range interactions
(\ref{gradient}). However, the regions 
occupied by one population need not be connected (cf. Figure~\ref{fig3}(B)).
If $S_{aa}< 0$ for all $a$, the distributions $\rho_a^{\rm st}(\vec{x})$ 
tend to be flat, as in the case of 
lane formation by repulsive pedestrian interactions
(Figure~\ref{fig3}(B)).
Instead, we have agglomeration (locally peaked stationary
solutions), if $S_{aa} >0$ for all $a$ (Figures~\ref{fig3}(C), (D)).
The example of trail formation, which is based on attractive
interactions between trails, corresponds to Figure~\ref{fig3}(C).
\par
Finally, we discuss the influence of noise. Adding diffusion terms
$\vec{\nabla} \cdot [D_a \vec{\nabla} \rho_a(\vec{x},t) ]$ to the right-hand
side of (\ref{cont}), the relation (\ref{endresult}) will not be
exactly valid anymore. Thus, optimality will be affected.
Moreover, the condition (\ref{cond}) for segregation will become
$\rho_a^{\rm hom} S_{aa} + \rho_b^{\rm hom} S_{bb} > 0$ or
$(S_{ab} S_{ba} - S_{aa} S_{bb}) \rho_a^{\rm hom}
 \rho_b^{\rm hom} > D_a D_b$.
Hence, growing diffusion coefficients will produce more homogeneous
equilibrium states, which agrees with intuition.
\par
We call attention to
the fact that there is a class of living systems (for which we
have given some realistic examples) to which existing
methods, notions, and principles of statistical mechanics 
can be successfully applied. In particular,
we have proven that systems, which can be represented as
a game between symmetrically interacting populations, approach
a stationary state characterised by maximal overall success and
minimal dissipation.
In other words, as individual entities are trying to optimise
their {\em own} success, a class of systems tends to reach a state with
the highest {\em global} success. This non-trivial result is valid
only under the conditions discussed above.
In contrast to self-organised optimality, self-optimisation without
a self-organised state occurs for symmetrical interactions,
if condition (\ref{cond}) is not fulfilled
(Figure~\ref{fig3}(A)). There
are also cases of self-organisation without optimality, if
(\ref{cond}) is satisfied, but the interactions are not symmetrical.
Such a system is exemplified by
uni-directional multi-lane traffic
of cars and lorries\cite{solid}, but even there
one observes significantly reduced interactions (lane-changing rates).
Furthermore, we point out that the above minimal dissipation principle
may be relevant to physical systems like driven granular 
media,\cite{granular} but the
most important implications of this generalised thermodynamic
concept are expected for 
biological, social, and economic systems.

{\em Acknowledgments:} 

D.H. wants to thank the DFG for financial support
by a Heisenberg scholarship.
This work was in part supported by OTKA F019299 and FKFP 0203/1997.
The authors are also grateful to Ill\'{e}s J. Farkas
for producing Figure~2.

\clearpage
\unitlength1.6cm
\clearpage
\unitlength10mm
\clearpage
\unitlength10mm
\end{document}